\documentclass{PoS}
\usepackage{subcaption}

\title{10 Years of XRT light curves: a general view of the X-ray afterglow}

\ShortTitle{10 Years of XRT light curves: a general view of the X-ray afterglow}

\author{\speaker{Onelda Bardho} \\
ARTEMIS (CNRS/OCA/UNS), Nice, France\\
E-mail: \email{onelda.bardho@oca.eu}}

\author{Michel Bo\"{e}r\\
 ARTEMIS (CNRS/OCA/UNS), Nice, France\\
	E-mail: \email{michel.boer@unice.fr}}
				
	\author{Bruce Gendre\\
	University of the Virgin Islands, Virgin Islands, USA\\
 ARTEMIS (CNRS/OCA/UNS), Nice, France\\
   E-mail: \email{bruce.gendre@gmail.com}} 

\abstract{During the pre-{\em Swift} era, a clustering of light curves was observed in the X-ray, optical and infrared afterglow of gamma-ray bursts. 
We used a sample of 254 GRB X-ray afterglows to check this fact in the {\em Swift} era. We corrected fluxes for distance, time dilation and losses of energy due to cosmological effects. 
With all our data in hand, we faced with a problem: our data were scattered. We investigated 3 possibilities to explain this, namely: the clustering does not exist, there are problems during calibration of data, and there are instrumental problems. 
We finally confirm that our sample is consistent with Dainotti correlation.}

\FullConference{Swift: 10 Years of Discovery,\\
		2-5 December 2014\\
		La Sapienza University, Rome, Italy }

\begin{document}

\section{Introduction}

Because of its long stay in orbit, {\em Swift} \cite{gehrels04} has observed numerous Gamma-Ray Bursts (GRBs) and has helped solving a lot of mysteries related to the GRBs research. 
Thanks to its fast capabilities and multi-wavelength instruments, {\em Swift} also opened a new era in the GRB science, and one can really refer to a pre-{\em Swift} era and a {\em Swift} era. In the field of standardization of GRBs, the most tricking results are that pre-{\em Swift} and {\em Swift} era results are sometime not consistent.

For instance, in the pre-{\em Swift} era, the light curves of GRB afterglows were known to cluster around defined paths once corrected for any cosmological and distance effects. This was firstly stated in X-ray by \cite{boer00}, and then confirmed by \cite{gendre04} and \cite{kou04}. A final (to date) check at that wavelength done by \cite{gendre08} by mixing pre-{\em Swift} and {\em Swift} light curves confirmed that fact: the light curves seem to cluster in two well organized groups and a third, more loosely distributed, faint group (see also Dereli et al., these proceedings, for this last group). Other groups also confirmed that fact in optical \cite{nardini06, liang06}, and infrared \cite{gendre_pelisson08}.

This changed in the {\em Swift} era. The profusion of X-ray light curves from the XRT allowed to study the distribution of the luminosity at a given time, and the conclusions of these works point toward no clustering in various groups, but instead in a loose and broad distribution (see for instance \cite{melandri14} for one of the latest of these works).

There is no obvious reason why the pre-{\em Swift} and {\em Swift} distributions of luminosities are not consistent. As a matter of consequence, we started a large project of re-analysis of a sample of X-ray afterglow light curves with known distance of the {\em Swift} era, in order first to confirm independently (and with the same method that was leading previously to the clusterings) this inconsistency; and second then to explain it.

Through all this paper, we will use a flat $\Lambda$CMD model of Universe, with $\Omega_m$ = 0.3. All errors are quoted at the 90 \% confidence level, and all methods are identical to those used in \cite{gendre08} for a fair comparison of the results.

\section{X--ray afterglow sample and data analysis}

We started to build our sample by adding to the sample of \cite{gendre08} the observations of long GRBs made by {\em Swift} between its launch and February 2013. We restricted this sample to bursts being observed by the XRT and having a measured redshift. This leads to a sample of 254 GRBs.

There are automated pipeline providing calibrated flux light curves \cite{evans07, evans09}. However, as stated before, we preferred to apply a similar method of the one of \cite{gendre08}, and thus we did not use these final light curves. We however do use the raw count light curves provided by the Leicester online repository. This is the only difference with the work done in \cite{gendre08}, as the manual extraction of more than 200 light curves would take too many time.

Once we have obtained these light curves and the associated event files, we apply our standard method, using the {\tt \small{HEASoft(v6.14)}} that includes {\tt \small{XSPEC,XIMAGE,XSELECT}} and  {\em Swift} data analysis tools. First we run the XRT tool {\tt \small xrtpipeline(v0.12.6)} to recalibrate the data using the latest available calibration. Being interested only in the afterglow part of the GRB, we removed from the data everything not related to the standard afterglow (i.e. the prompt phase when visible, the plateau phase, and the flares). We then extract a spectrum from the remaining data, and fit it with absorbed power law model ({\tt \small{wabs*zwabs*pow}}). One of the absorption component is fixed to the galactic value, while the second is let free to vary at the redshift of the GRB. The best fit model allows us to compute the Energy Correction Factor (ECF) between counts and flux, in order to convert the raw count light curves in the standard 2-10 keV flux light curves.
 
To be consistent with previous papers we corrected the distance by normalizing all light curves in the standard distance to a redshift of $z = 1$. We thus correct for time dilation and we worked on 'rescaled flux' instead of luminosity. This is not mandatory to see the clusterings (\cite{liang06} were working on luminosity), but allows for a more precise k-correction of the light curves, and thus add less systematic dispersion into any distribution. We finally build a normalized flux distribution at one day after the burst (in the common $z = 1$ frame) from these light curves.

\section{Results}

\begin{figure}
\centering
\includegraphics[width=0.45\linewidth]{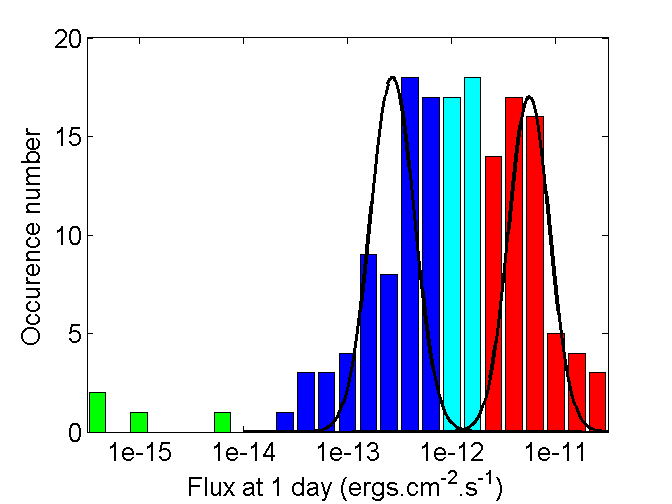}
\includegraphics[width=0.45\linewidth]{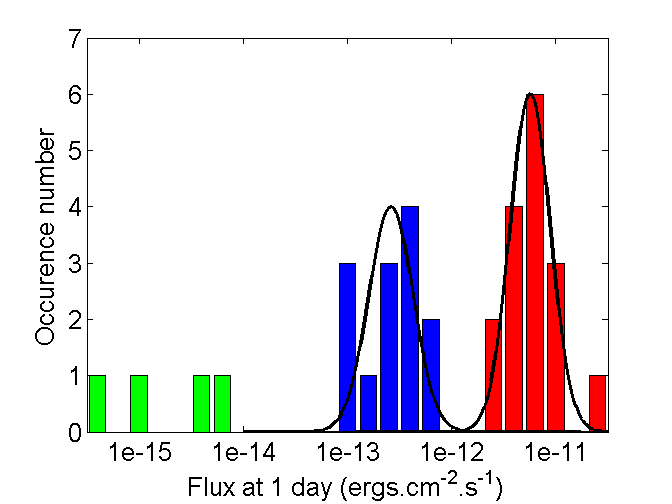}
\caption{Flux distribution at 1 day after the burst. Left: our sample. Right: for comparison, the sample of \cite{gendre08}. Red, blue, and green bars show groups I, II, and III respectively. The cyan bars show the GRBs between Groups I and II. The black line shows the best-fit Gaussian distribution of \cite{gendre08}.\label{fig1}}
%\label{fig:test}
\end{figure}

The final distribution we obtain is presented in Fig. \ref{fig1} (left). For comparison, we also inserted the results of \cite{gendre08} as a similar distribution (right part of the figure). We can clearly see that, in agreement with the previous {\em Swift} era results, there is no clear clustering. In fact, we do observe 35 GRBs located between the groups I and II.

\begin{figure}
\centering 
\includegraphics[scale=0.4]{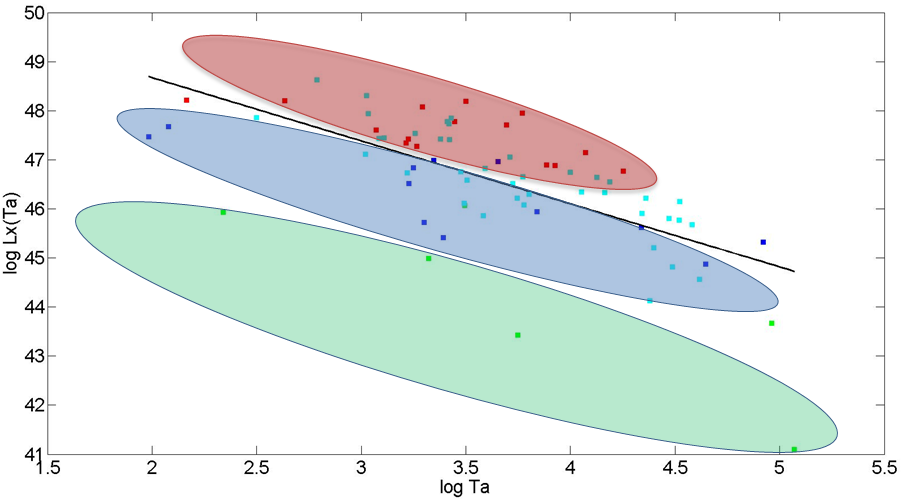}  
\caption {Dainotti correlation log($L_X(T_a)$) versus log($T_a$). Red, blue and green points show Groups I, II and III respectively. 
Cyan points are 35 GRBs that are between Groups I and II.}
\label{fig:dainotti_bruce_plus_middle_data}
\end{figure}

In order to check for any processing problems, we also checked a known {\em Swift} era correlation, i.e. the Dainotti correlation \cite{dainotti08, dainotti10}, that links the date of the end of the plateau phase and the luminosity at that time. This correlation has its properties related to the clustering of light curves. The Fig. \ref{fig:dainotti_bruce_plus_middle_data} present our check: we can reproduce the correlation, with its normal intrinsic distribution. We can also note that the red and blue points are distributed in two parallel layers with nearly no "systematic noise" within each group, which could indicate that the Dainotti correlation is stronger when considered for only one group of events. However, the 35 GRBs located between groups I and II (cyan points) are still problematic since they are placed everywhere and do not fit in any group. \cite{dainotti08, dainotti10} have worked only with GRBs that show a plateau phase, while we worked with all GRBs, considering in this case an extended version of Dainotti correlation. This should not impact our results because we are working only with the beginning of afterglow, that is considered as the end of plateau phase \cite{willingale_07}.

\section{Discussion}

\begin{figure}
\centering 
\includegraphics[width=0.45\textwidth]{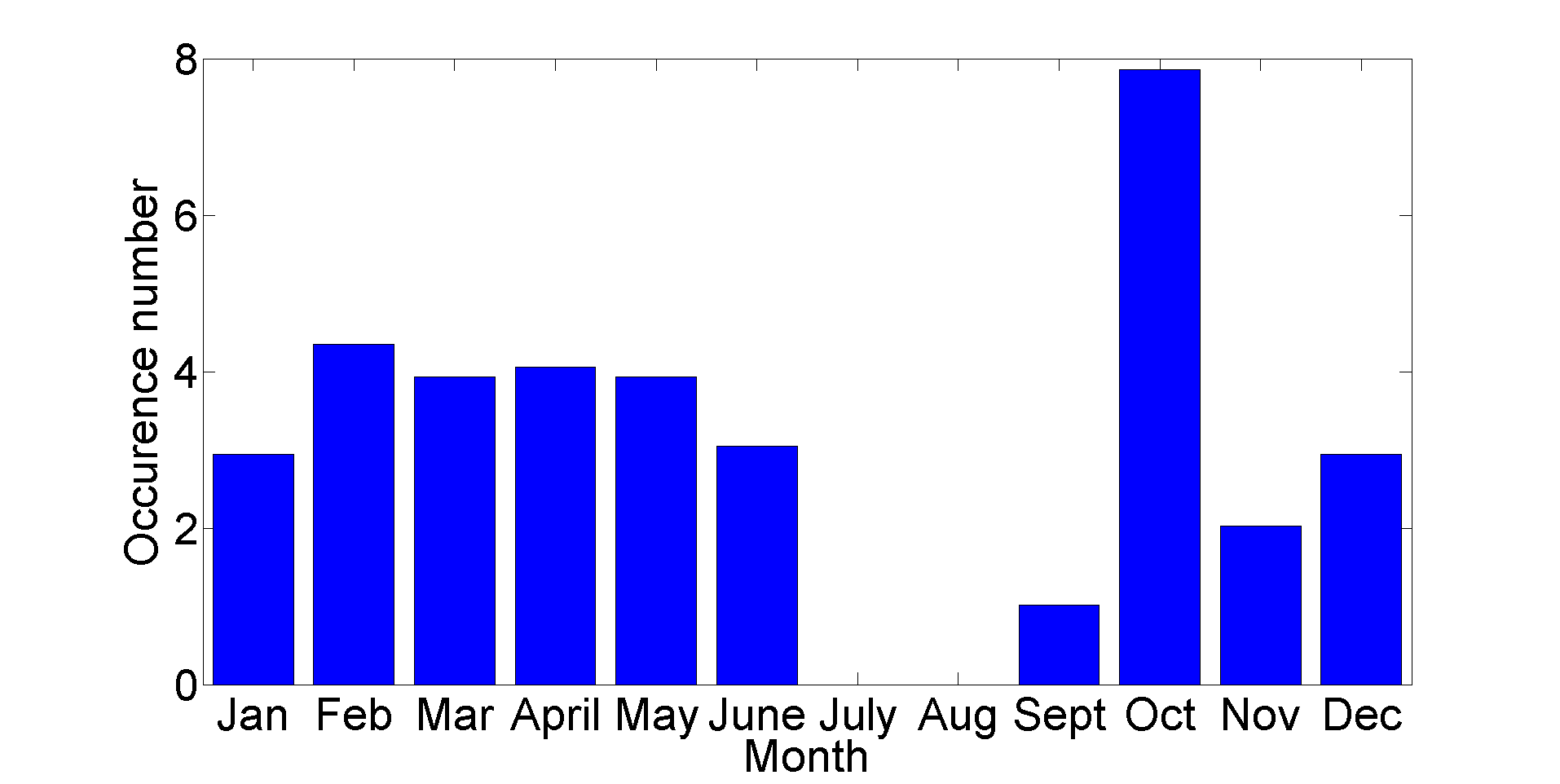}  
\includegraphics[width=0.45\textwidth]{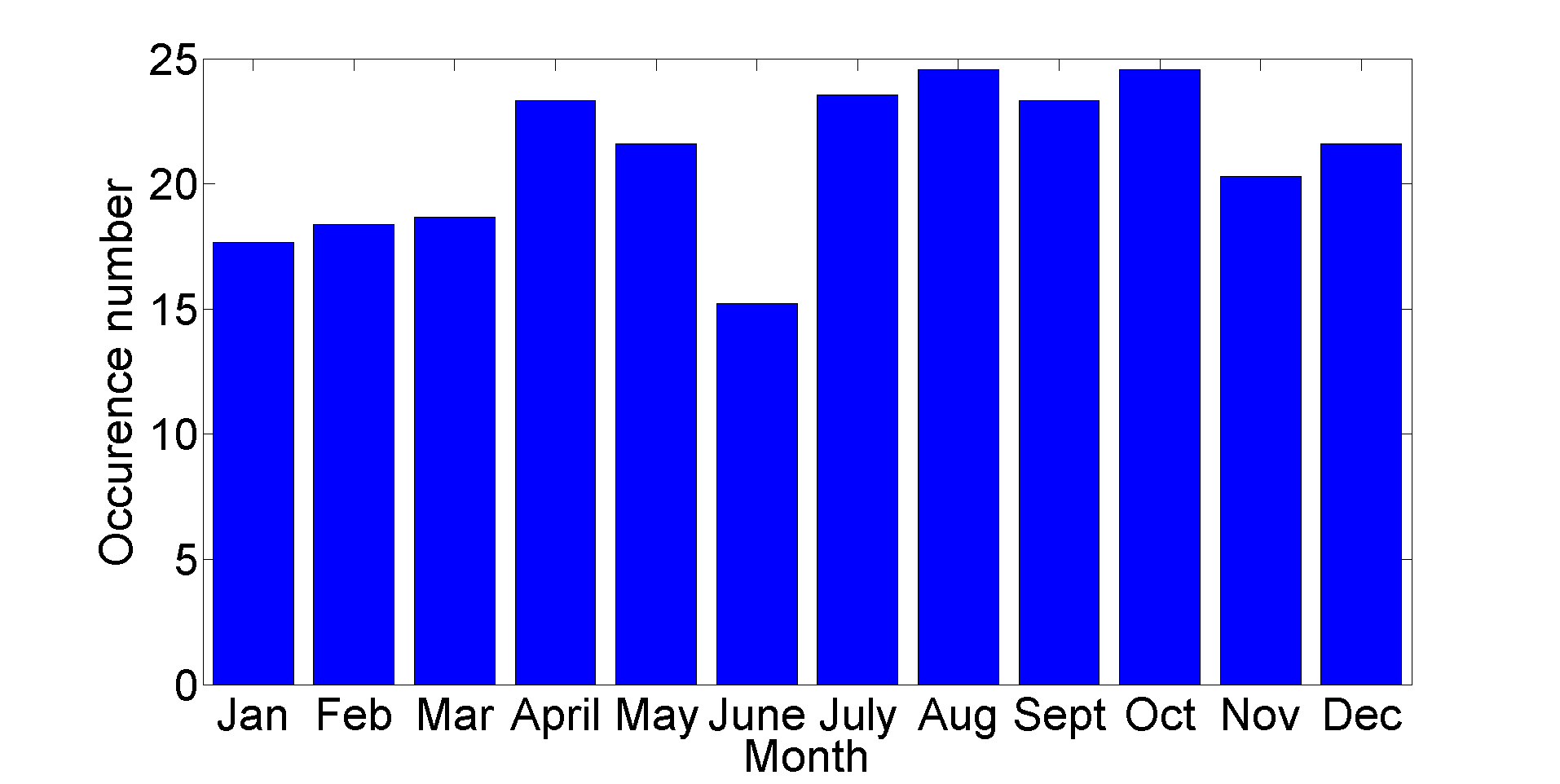}  
\caption {Left: Distribution of the month of detection for the 35 GRBs located between groups I and II. Right: Same distribution, but for our whole sample.\label{fig3}}
\end{figure}

In the previous section, we have shown that in the {\em Swift} era, even when using our (slightly modified) method, we do not observe anymore the clustering of light curves. This leads to an interesting problem: why this clustering was observed in the pre-{\em Swift} era and not now? And does this also occurs for other GRB properties?

The obvious answer to the first question would be a selection effect. In the pre-{\em Swift} era, the rarity of X-ray light curves implies to build a sample of events observed by BeppoSAX, XMM-Newton, and Chandra. It is this mix that produce an (apparent) lack of events, then solved by {\em Swift}. This argument however does not hold, as the clustering was also observed when using only BeppoSAX data \cite{boer00}.

The fact that only the addition of {\it some} (not all) {\em Swift} bursts destruct the clustering made us wondering if there is not an instrumental problem. We checked the date of observation of these problematic events, and found some kind of periodicity (see Fig. \ref{fig3}, left part), with less problems near the solstices, and a maximum of problematic events near the equinoxes. This is outlined when one compare with the same distribution, but for all bursts (presented in Fig. \ref{fig3}, right part), where there is absolutely no periodicity (one should note that we normalized each month by its duration, so February and January can be fairly compared).

The first idea one can have at that time is a problem of Earth (or Sun) limb, i.e. an increase of the background induced by the orbit, and maybe not well taken into account during the data reduction. This could affect the raw light curve, or the spectrum. We thus tried to study both hypotheses.

\begin{figure}
\centering 
\includegraphics[scale=0.2]{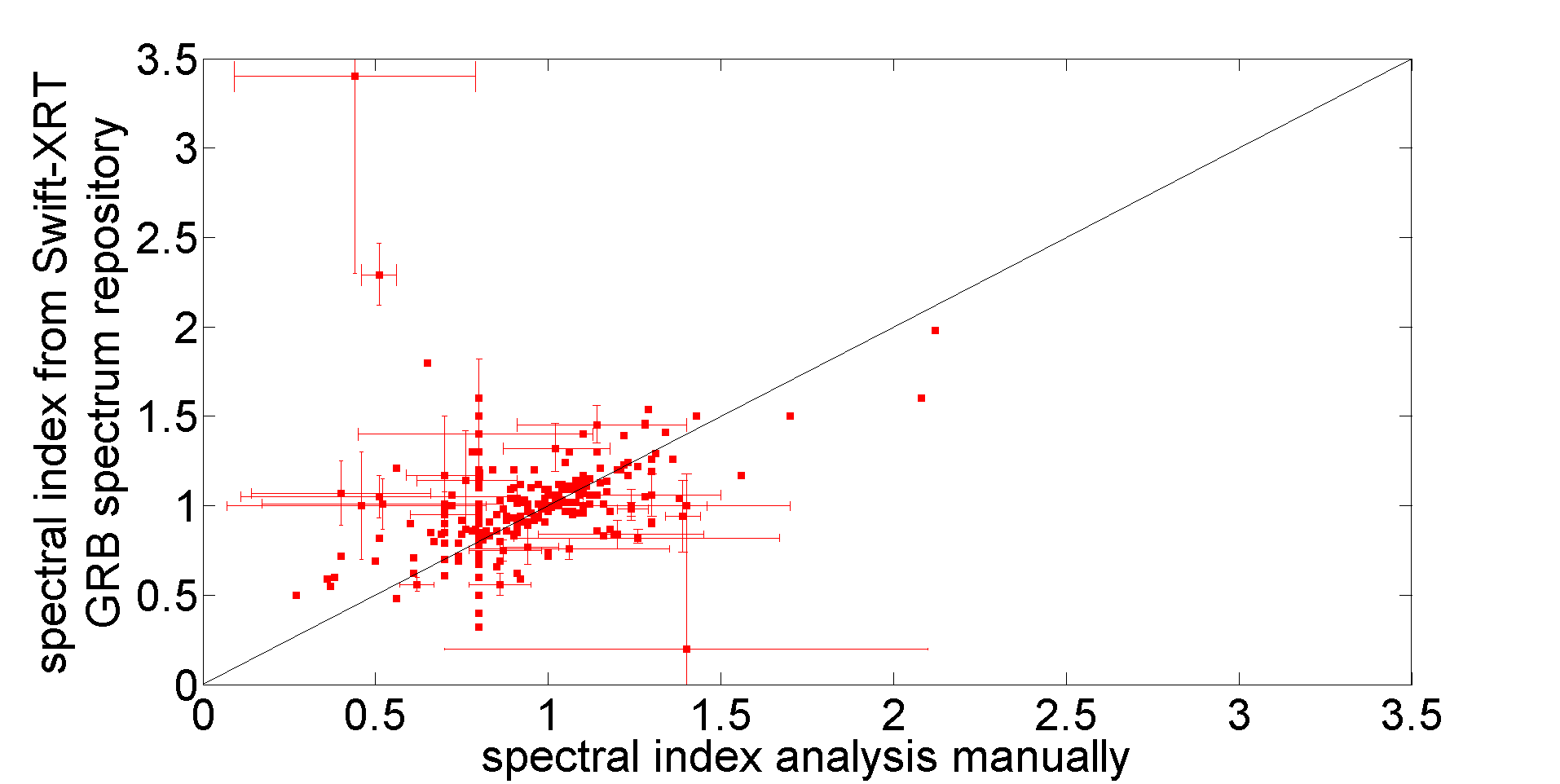}  
\caption {Comparison of the spectral indices from the manual analysis (x axis) and {\it Swift}-XRT GRB spectrum repository (y axis). The black line is the line of equality, where both results are equal. We removed for clarity from the plot the error bars of GRBs which are compatible with equality.}
\label{fig:spectral_index_comparision_paper}
\end{figure}

We compared the spectral index of our manual analysis and the results from the {\em Swift}-XRT GRB spectrum repository \cite{evans07, evans09}. The result is displayed in Fig. \ref{fig:spectral_index_comparision_paper}. Clearly, in most of the cases, we are in close agreement. However, in about 11.2 \% of the case, we found some discrepancies. We stress that this comparison is not straightforward, as we are not dealing with the exact same thing. First, in some cases, the temporal range for the extraction of the spectrum is not the same. We have extracted the spectrum of the {\em afterglow}, while the {\em Swift}-XRT GRB spectrum repository propose by default a time averaged spectrum between T$_1$ and T$_2$ of the object observed, irrespectively of its emission phase\footnote{A customized analysis, selecting the times T$_1$ and T$_2$ at the user requests, is however possible within online repository}. 
Second, each fit is done by using an absorbed power law model ({\tt \small{wabs*zwabs*pow}}) one fixed in our own galaxy and one let free to the redshift, while the {\em Swift}-XRT GRB spectrum repository is using the same model but with two small differences: a. For some GRBs, the spectrum repository webpage reports an unknown value for the redshift, and in that case the spectral fit can give inconsistent results and b. {\em Swift}-XRT GRB spectrum repository is using {\tt \small phabs} as an absorption model while we are using {\tt \small wabs}. Last, the methods of correcting for pile-up are very different and can lead to strong differences. We are using the method explained at \cite{vaughan_06}, while the limit for considering a pile-up region it is 0.8 counts/sec, but the {\em Swift}-XRT GRB spectrum repository is using 0.6 counts/sec \cite{evans07, evans09}.
We thus do expect some differences, and found that about 90 \% of compatibility is a good result.
There are in any case 28 bursts with significant spectral discrepancies. Unfortunately, these are not all in the group of 35 problematic events.
This is not the first main explanation so we are working in light curves to understand it better.

\section{Conclusions}

We have constructed a sample of 254 X-ray afterglows of {\em Swift} GRBs in order to check the clustering of afterglow light curves observed during the pre-{\em Swift} era. We can clearly see that in the {\em Swift} era, the clustering is no more apparent. We observe the presence of 35 bursts at the place of the separation of the two brightest groups. We have checked our analysis for a processing problem by reproducing the Dainotti correlation, found in the Swift era, and we successfully observed the correlation. We noted that each clustering group is consistent with the Dainotti correlation, and that could mean a kind of relationship between with correlation and the clustering correlation.

We also checked for instrumental effects and found that we have some seasonal effects for the 35 "between groups" GRBs. This could indicate a problem of background induced by the orbit. Checking our spectral analysis with the one of the {\em Swift}-XRT GRB spectrum repository, we found that in the majority of cases we are in agreement. However, for 11.20 \% of bursts, we found some discrepancies. Some can be easily explained by a difference of time range (for the extraction of the spectrum) or model, while others are more difficult to understand. We are still working on this to understand the origin of this discrepancy.

\section{Acknowledgement}

OB is supported by the Erasmus Mundus Joint Doctorate Program by Grant Number 2012-1710 from the EACEA of the European Commission. This work made use of data supplied by the UK Swift Science Data Center at the University of Leicester.

\end{document}